\documentclass[a4paper,11pt]{article}
\pdfoutput=1 

\usepackage{jheppub} 

\newcommand {\bc}{\begin{center}}
\newcommand {\ec}{\end{center}}
\newcommand {\bea}{\begin{eqnarray}}
\newcommand {\eea}{\end{eqnarray}}
\newcommand {\be}{\begin{equation}}
\newcommand {\ee}{\end{equation}}

\def\lsim{\mathrel{\rlap{\lower4pt\hbox{$\sim$}}
    \raise1pt\hbox{$<$}}}               
\def\gsim{\mathrel{\rlap{\lower4pt\hbox{$\sim$}}
    \raise1pt\hbox{$>$}}}  
\usepackage{graphicx}
\usepackage{amsmath}

\begin{document}


\title{N-particle irreducible actions for stochastic fluids}

\author[a]{Jingyi~Chao}
\author[b]{Thomas~Sch\"afer}

\affiliation[a]{College of Physics and Communication Electronics, 
Jiangxi Normal University, Nanchang 330022, China}
\affiliation[b]{Department of Physics, North Carolina State University,  
Raleigh, NC 27695}

\emailAdd{chaojingyi@jxnu.edu.cn}
\emailAdd{tmschaef@ncsu.edu}
\arxivnumber{2302.00720}

\abstract{
 We construct one- and two-particle irreducible (1PI and 2PI)
effective actions for the stochastic fluid dynamics of a
conserved density undergoing diffusive motion. We compute 
the 1PI action in one-loop order and the 2PI action in two-loop
approximation. We derive a set of Schwinger-Dyson equations 
and regularize the resulting equations using Pauli-Villars 
fields. We numerically solve the Schwinger-Dyson
equations for a non-critical fluid. We find that higher-loop
effects summed by the Schwinger-Dyson renormalize the non-linear 
coupling. We also find indications of a diffuson-cascade, the 
appearance of $n$-loop correction with smaller and smaller
exponential suppression.}

\maketitle

\section{Introduction}
\label{sec_intro}

  The study of hydrodynamic fluctuations has received renewed interest
in connection with the experimental search for a conjectured critical 
endpoint in the phase diagram of Quantum Chromodynamics (QCD) \cite{
Stephanov:1998dy,Bzdak:2019pkr,Bluhm:2020mpc,An:2021wof}. The basic 
idea is that the quark-gluon plasma created in a heavy ion collision
is a locally equilibrated fluid, and that each fluid element traces
out a trajectory in the QCD phase diagram. If the trajectory approaches
a critical point then the correlation length will grow, and fluctuations
of thermodynamic variables are enhanced. At freezeout fluctuations in
the fluid are converted to fluctuations of particle distributions, which
can be measured experimentally. A typical set of observables is given
by the cumulants of the net-proton number in a given rapidity window.  

 In thermodynamic equilibrium fluctuations in conserved densities are 
governed by susceptibilities, which can be obtained as derivatives of 
the thermodynamic potential. The theory of second-order phase transitions
predicts that near the critical point susceptibilities scale as powers
of the correlation length $\xi$, where higher-order susceptibilities 
scale with a larger power of $\xi$. Higher-order susceptibilities also 
potentially exhibit an oscillatory dependence on control parameters, such 
as the temperature and the baryon chemical potential. Both of these
observations imply that non-Gaussian cumulants provide crucially 
consistency checks for the possible discovery of a critical endpoint 
\cite{Ejiri:2005wq,Stephanov:2008qz,Asakawa:2009aj,Stephanov:2011pb,
Friman:2011pf}.

  The fluid created in a heavy ion collision expands rapidly, and 
fluctuation observables are expected to deviate from equilibrium 
expectations. In the vicinity of a critical point non-equilibrium 
phenomena, such as critical slowing down cannot be ignored 
\cite{Berdnikov:1999ph,Nahrgang:2018afz,Akamatsu:2018vjr,Bluhm:2020mpc,
An:2021wof}. Dynamical critical scaling predicts the dependence of 
the relaxation time $\tau$ on the correlation length, $\tau\sim \xi^z$,
where $z$ is the dynamical critical exponent \cite{Hohenberg:1977ym}.
While the value of $z$ for theories in a different universality 
classes are known from numerical calculations and the epsilon 
expansion \cite{Hohenberg:1977ym,Folk:2006ve}, less is known about
the functional form of time-dependent $n$-point functions, and the
relative relaxation rate of $n$-point functions for different $n$. 

 Several methods for studying hydrodynamic $n$-point functions have
been explored in the literature. This includes numerical simulations
of stochastic fluid dynamics \cite{Berges:2009jz,Nahrgang:2018afz,
Schweitzer:2020noq,Schweitzer:2021iqk,Pihan:2022xcl,Schaefer:2022bfm,
Chattopadhyay:2023jfm}, dynamical evolution equations for $n$-point 
functions \cite{Mukherjee:2015swa,Akamatsu:2016llw,Stephanov:2017ghc,
Martinez:2018wia,Akamatsu:2018vjr,An:2019osr,An:2019csj,An:2020vri},
as well as hydrodynamic effective actions \cite{Liu:2018kfw,
Chen-Lin:2018kfl,Chao:2020kcf,Sogabe:2021svv}. In previous work we
considered a pure perturbative approach to effective actions for 
fluid dynamics \cite{Chao:2020kcf}. In the present paper we study 
a non-perturbative approach based on $n$-particle irreducible
effective actions, see \cite{Berges:2004yj,Calzetta:2008iqa} for 
a review. In the following we develop the formalism in the context
of a simple model of non-linear diffusion, and we study a numerical 
solution of the Schwinger-Dyson equation for the two-point function
in a non-critical theory. 

\section{1PI effective action}
 
  We consider a conserved density $\psi(x,t)$. In thermal equilibrium
the probability distribution of $\psi$ is governed by a free energy
functional
\be
\label{F-modB}
{\cal F}[\psi]  = \int d^3x \, \left\{ 
 \frac{1}{2} (\vec\nabla \psi)^2
 +  \frac{m^2}{2}\, \psi(x,t)^2 +\frac{\lambda_3}{3!}\, \psi(x,t)^3 
 +  \dots  \right\} \, .
\ee
where $\ldots$ contains higher-order non-linearities. Of course, 
for the theory to be stable there has to be a fourth-order (or
higher order even) interaction present as well. In the vicinity 
of a critical point in the Ising universality class there is an 
emergent $Z_2$ symmetry and the cubic term is absent, but at a 
generic point in the phase diagram a cubic non-linearity is 
present. The dynamics of the theory are governed by a 
diffusion equation 
\be
\partial_t \psi(x,t) = \kappa\nabla^2 
  \left( \frac{\delta {\cal F}[\psi]}{\delta\psi(x,t)}\right)
   + \theta(x,t), 
\ee
where $\kappa$ is a conductivity, $D=\kappa m^2$ is the 
diffusion constant, and $\theta(x,t)$ is a noise term. 
The noise has zero mean $\langle\theta(x,t)\rangle=0$
and correlation 
\be 
\langle \theta(x,t)\theta(x',t')\rangle = 
  -\kappa T\nabla^2 \delta(x-x')\delta(t-t').
\ee
The structure of the noise correlator is fixed by 
fluctuation-dissipation relations, and ensures that the 
equilibrium distribution is given by $P[\psi]\sim\exp(-{\cal F}
[\psi]/T)$.

 The observables of the theory are correlation functions 
of the density $\psi(x,t)$ averaged over different realizations
of the noise. It is well known that these correlation functions
can be derived from an effective lagrangian that contains an
additional auxiliary field $\tilde\psi$ \cite{Martin:1973zz,
Janssen:1976,DeDominicis:1977fw,Chen-Lin:2018kfl,Liu:2018kfw}. 
The effective lagrangian is given by 
\be
{\cal L} = \tilde\psi \left(
  \partial_t - D\nabla^2 \right)\psi
  + \kappa T\tilde\psi\nabla^2\tilde\psi
  - \frac{\kappa\lambda_3}{2}\; \tilde\psi \nabla^2 \psi^2 
  + \ldots \, .
\ee
Note that the variation of the action with respect to $\tilde\psi$ 
leads to a stochastic diffusion equation for $\psi$, where the 
structure of the noise term is governed by the quadratic 
term in $\tilde\psi$. We consider the partition function
\be
\label{Z_J}
 Z[J,\tilde{J}] = \int D\psi D\tilde\psi\, 
    \exp(-S)\, , \hspace{1cm}
    S=\int dt\, d^3x \left\{ {\cal L} + 
        \tilde{J}\tilde\psi + J\psi\right\}\, . 
\ee
We define $\exp(-W)=Z$. Then
\be 
\label{W_J}
 \frac{\delta W}{\delta J} = \langle \psi\rangle = \Psi\, , 
 \hspace{1cm}
 \frac{\delta W}{\delta \tilde{J}} 
    = \langle \tilde\psi\rangle = \tilde\Psi\, , 
\ee
and we can define the Legendre transform
\be
\label{Gamma_Psi}
\Gamma[\Psi,\tilde\Psi] = W[J,\tilde{J}]
  - \int dt\, d^3x \left( J\Psi + \tilde{J}\tilde\Psi \right)\, . 
\ee
This relation defined the 1PI effective action $\Gamma$. We 
can compute the effective action using the background field
method. We write
\be 
\label{psi_cl_qu}
 \psi = \Psi+ \delta \psi \, , \hspace{1cm}
 \tilde\psi = \tilde\Psi + \delta\tilde\psi \, . 
\ee
Then
\be
\label{S_BF}
S[\psi,\tilde\psi]  = S[\Psi,\tilde\Psi] + 
  \int dt\,d^3x\,\left( \frac{\delta S}{\delta \Psi}\delta\psi
     + \frac{\delta S}{\delta \tilde\Psi}\delta\tilde\psi\right)
     + S_2 [\delta\psi,\delta\tilde\psi,\Psi,\tilde\Psi] \, , 
\ee
where the fluctuation term $S_2$ is given by 
\bea 
\label{S_2}
S_2 [\delta\psi,\delta\tilde\psi,\Psi,\tilde\Psi] 
  &=& \int dt\,d^3x\,  \Big\{
\delta\tilde\psi \left(
  \partial_t - D\nabla^2 \right)\delta\psi
  + \kappa T\delta\tilde\psi\nabla^2\delta\tilde\psi 
  - \frac{\kappa\lambda_3}{2}\; \delta\tilde\psi \nabla^2 
     (\delta\psi)^2 \nonumber \\
&&  \mbox{} -
  \frac{\kappa\lambda_3}{2}\; 
   \left[  (\nabla^2\tilde\Psi) (\delta\psi)^2 
      +  2 (\nabla^2\delta\tilde\psi)(\delta\psi)\Psi\right]
      \Big\}  \, . 
\eea
\begin{figure}[t]
\includegraphics[width=0.99\textwidth]{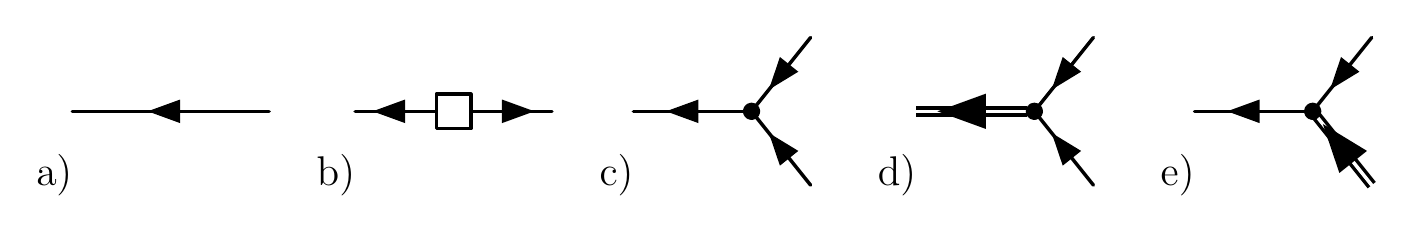}
\caption{\label{fig:Feyn-1PI}
Feynman rules for the calculation of the 1PI effective action.
Fig.~a) shows the propagator $\langle\delta\psi\delta\tilde\psi
\rangle$, b) is the propagator $\langle\delta\psi\delta
\psi\rangle$, c) is the non-linear $\delta\tilde\psi
(\delta\psi)^2$ interaction, and d,e) are the coupling to the 
classical fields $\tilde\Psi$ and $\Psi$, respectively. }
\end{figure}
We can now perform the Legendre transform. We find
\be 
 \Gamma[\Psi,\tilde\Psi] = S[\Psi,\tilde\Psi]
   + \Gamma_F[\Psi,\tilde\Psi] \, , 
\ee
where $\Gamma_F$, the fluctuation term, is 
\be 
 \Gamma_F[\Psi,\tilde\Psi] = \int D(\delta\psi\delta\tilde\psi)\, 
  \exp\left(-S_2[\delta\psi,\delta\tilde\psi,\Psi,\tilde\Psi]\right)\, . 
\ee
The 1PI effective action is given by the classical action 
$S[\Psi,\tilde\Psi]$ and fluctuation corrections generated by 
$S_2$. These corrections can be computed perturbatively, using 
the Feynman rules shown in Fig.~\ref{fig:Feyn-1PI}. The Feynman
rules for the fluctuating fields are identical to those given 
in our earlier work, see equ.~(3.1-3.3) in \cite{Chao:2020kcf}.
The vertices for $\Psi$ and $\tilde\Psi$ are new, and follow
by substitution from the cubic vertex $\tilde\psi\nabla^2
(\psi^2)$.

\begin{figure}[t]
\includegraphics[width=0.32\textwidth]{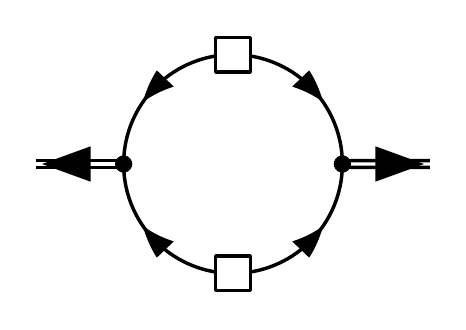}
\includegraphics[width=0.32\textwidth]{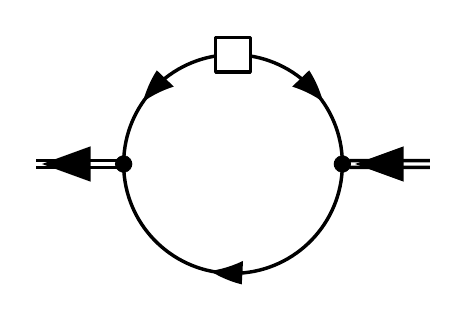}
\includegraphics[width=0.32\textwidth]{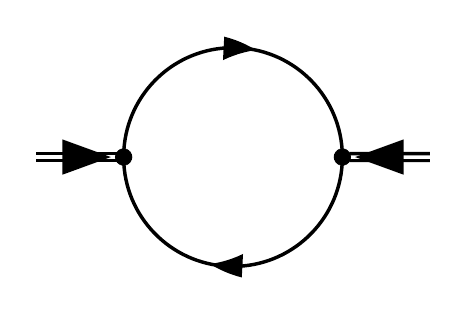}
\caption{\label{fig:1PI-loop}
One-loop contributions to the 1PI effective action. The left panel
shows the $\tilde\Psi^2$ term in equ.~(\ref{Gamma_1loop}), and the 
middle panel is the $\tilde\Psi\Psi$ term. The right panel shows
the $\Psi^2$ term, which vanishes.}
\end{figure}

 We expand $\Gamma_F$ in powers of $\Psi$ and $\tilde\Psi$. 
Linear terms correspond to tadpole diagrams, which vanish. One-loop 
contributions to the quadratic terms are shown in 
Fig.~\ref{fig:1PI-loop}. Higher loop corrections are suppressed 
by powers of the external momentum. The one-loop terms are 
\bea
\label{Gamma_1loop}
\Gamma_F[\Psi,\tilde\Psi] &=&  \int dt\,dt'\,d^3x\, d^3x'\,
  \tilde\Psi(x,t)\Psi(x',t')\Sigma(x-x',t-t') \nonumber \\
 && \mbox{} +  \int dt\,dt'\,d^3x\, d^3x'\,
  \tilde\Psi(x,t)\tilde\Psi(x',t')\delta D(x-x',t-t')\, . 
\eea
In a large homogeneous system the self energies $\Sigma$ and 
$\delta D$ can be computed in frequency-momentum $(\omega,k)$ space. 
Using the results from \cite{Chao:2020kcf,Chen-Lin:2018kfl} we have
\bea 
\label{Sigma_1PI}
 \Sigma(\omega,k) &=& \frac{i\lambda_3^2 T}{32\pi m^6}\,
  \omega k^2\sqrt{k^2-\frac{2i\omega}{D}} \, , \\
\label{D_1PI}
 \delta D(\omega,k) &=& \frac{D\lambda_3^2 T^2}{16\pi m^8}\,
   k^4 {\it Re} \sqrt{k^2-\frac{2i\omega}{D}}\, . 
\eea
We can now determine equations of motion for $(\Psi,\tilde\Psi)$ 
that include the fluctuation effects encoded in 
equ.~(\ref{Sigma_1PI}, \ref{D_1PI}). In particular, there is a 
"classical solution" with $\tilde\Psi=0$ and
\be 
\label{Psi_EoM}
(\partial_t -D\nabla^2)\Psi 
 - \frac{\kappa\lambda_3^2}{2}\nabla^2 \Psi^2
  + \int d^3x'\, dt'\, \Psi(x',t')
    \Sigma(x,t;x',t') = 0\, . 
\ee
Here, the first term is the classical diffusion equation, the
second term encodes non-linearities, and the third term 
accounts for fluctuation effects. To understand these effects,
consider a mixed representation $\Psi_k(t)$, where we have 
performed a Fourier transform with respect to the spatial 
coordinate. The classical diffusion equation corresponds 
to an exponential decay, $\Psi_k(t)\sim \exp(-Dk^2t)\Psi_k(0)$.
Fluctuation effects are described by the mixed-representation
self energy $\Sigma(t,k)$. Using the one-loop result in 
equ.~(\ref{Sigma_1PI}) we get 
\be 
\label{sigR_t_k}
\Sigma(t,k) = \frac{3(\kappa\lambda_3)^2}{16\pi^{3/2}} 
    \, \frac{Tk^2}{m^2}\,
   \Theta(t) \exp\left(-\frac{Dk^2 t}{2}\right)
    \left\{ \frac{1}{(2Dt)^{5/2}} + \frac{k^2}{6(2Dt)^{3/2}}\right\},
\ee
where we have performed the Fourier transform using contour 
integration. The fractional powers of $t$ characterize a long time 
tails in the evolution of the density due to fluctuations.
The correction to the noise term is given by
\be 
\label{delD_t_k}
\delta D(t,k) = \frac{(\kappa\lambda_3)^2}{32\pi^{3/2}}
  \, \frac{T^2}{m^4} \, \frac{k^4}{(2D|t|)^{3/2}}
  \, \exp\left( -\frac{Dk^2|t|}{2} \right) \, . 
\ee
This result shows that the effective noise term is non-local
in both space and time. 

\section{2PI effective action}

\begin{figure}[t]
\begin{center}
\includegraphics[width=0.23\textwidth]{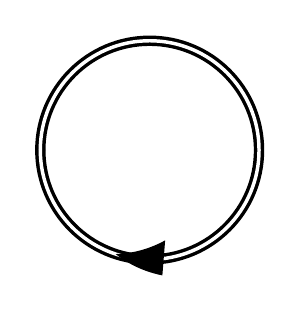}
\hspace*{0.02\hsize}
\includegraphics[width=0.23\textwidth]{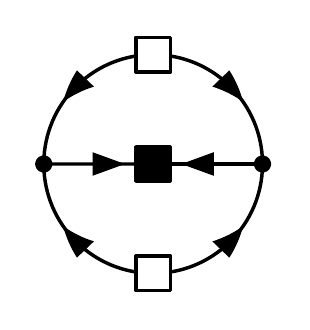}
\hspace*{0.02\hsize}
\includegraphics[width=0.23\textwidth]{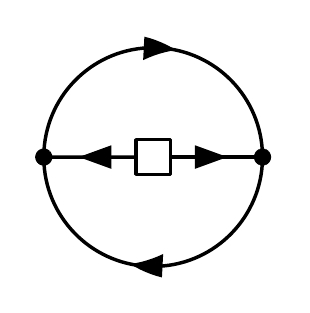}
\end{center}
\caption{\label{fig:2PI-loop}
One and two-loop contributions to the 2PI effective action. Here, 
the left panel corresponds to ${\rm Tr}\log G$, and the middle and
the right panel are two-loop diagrams constructed from the vertex in 
$S_3$ and the propagators $G_{12}$, $G_{21}$ (solid lines), 
$G_{22}$ (line with open box), and $G_{11}$ (line with solid box).}
\end{figure}

 We observed that the 1PI effective action is a convenient tool
for deriving an equation of motion that takes into account long
time tails in the evolution of the hydrodynamic field. This suggests
that we can use higher nPI effective actions to derive similar 
equations of motion for higher n-point functions. In order to
study 2PI and 3PI effective actions we will use a two-component 
notation $\psi_a=(\tilde\psi,\psi)$ for the hydrodynamic field. 
The quadratic action is 
\be 
 S = \frac{1}{2} \int d^3x\, dt\, \psi_a A_{ab} \psi_b\, , 
\ee
with 
\be 
A = \left( \begin{array}{cc}
  2\kappa T\nabla^2  &  \partial_t - D\nabla^2 \\
-\partial_t - D\nabla^2 & 0    
 \end{array}\right) \, . 
\ee
In $(\omega,k)$ space this leads to the matrix propagator
\be 
 G^0_{ab} = \frac{1}{\omega^2+(Dk^2)^2}
  \left( \begin{array}{cc}
  0 &  -i\omega + Dk^2 \\
  i\omega + Dk^2 & 2\kappa T k^2
  \end{array}\right) \, . 
\ee
It is interesting to note that this propagator has the 
analytical structure of the Closed Time Path (CTP) propagator
in the Keldysh basis, see, for example~\cite{Kamenev:2009jj}.
The interaction term can be written as 
\be
\label{S_I_K}
S = -\int d^3x\, dt\,  \frac{\kappa\lambda_3}{2}
    c_{abc} (\nabla^2\psi_a)\psi_b\psi_c\, , 
\ee
with $c_{122}=1$ and all others $c_{abc}=0$. Note that other 
structures are also possible. As explained in \cite{Chao:2020kcf}
$T$-reversal invariance is consistent with a coupling of the 
form $c_{abc}\psi_a(\nabla\psi_b)(\nabla\psi_c)$ with $c_{211}=1$
and all others $c_{abc}=0$. In addition to the local source term 
in equ.~(\ref{Z_J}) we also couple a bi-local source $\frac{1}{2}
\psi_a K_{ab}\psi_b$. Then 
\bea 
\label{W_J_a}
 \frac{\delta W}{\delta J_a} 
   &=& \langle \psi_a \rangle  = \Psi_a \, , \\
\label{K_J_ab}
 \frac{\delta W}{\delta K_{ab}} 
   &=& \frac{1}{2} \langle \psi_a\psi_b \rangle  
     = \frac{1}{2} \left[ \Psi_a\Psi_b + G_{ab} \right]\, , 
\eea 
where $G_{ab}$ is the full two-point function. We can now
perform a Legendre transform 
\be 
\label{Gam_2PI}
\Gamma[\Psi_a,G_{ab}] = W[J_a,K_{ab}]
 - J_A\Psi_A - \frac{1}{2}K_{AB}
    \left[ \Psi_A\Psi_B  + G_{AB} \right]\, , 
\ee
where we have introduced the notation $F_AG_A = \int d^3x\, dt\,
F_aG_a$. The 2PI effective action can be computed in analogy 
to 1PI action, using the background field method, see 
equ.~(\ref{psi_cl_qu}). We obtain
\be 
\label{Gam_2PI_BF}
\Gamma[\Psi_a,G_{ab}] = S[\Psi_a] 
 + \frac{1}{2}\,\frac{\delta^{2} S}{\delta \Psi_A\delta \Psi_B}\, G_{AB}
 - \frac{1}{2}\, {\rm Tr} \left[ \log (G)\right]
  + \Gamma_F[\Psi_a,G_{ab}]\, , 
\ee
where $G=\det G_{ab}$. The first term in equ.~(\ref{Gam_2PI_BF}) is 
the classical action, and the third term is the one-loop correction
generated by the full propagator $G$. The second term ensures that 
the leading term in the equation of motion for the propagator is 
$G=G^0$. Higher order fluctuations are described by $\Gamma_F$,
\bea 
\label{Gam_2PI_int}
 \exp(-\Gamma_F[\Psi_a,G_{ab}] ) &=& \frac{1}{\sqrt{\det(G)}}
   \, \int D(\delta\psi_a) \, 
   \exp\Big\{ -\frac{1}{2} \delta\psi_A (G^{-1})_{AB} \delta\psi_B
   \nonumber \\
    && \hspace*{1cm}\mbox{}
     - \left[ 
     S_3[\Psi_a,\delta\psi_a] - \bar{J}_A\delta\psi_A
       -\bar{K}_{AB} (\delta\psi_A\delta\psi_B-G_{AB}) \right]
       \Big\},
\eea 
where we have introduced \cite{Calzetta:2008iqa}
\be
\label{Gam_2PI_sources}
\bar{J}_a = \frac{1}{2} \,
    \frac{\delta^3 S}{\delta\Psi_a\delta\Psi_B\delta\Psi_C}\, G_{BC}
     + \frac{\delta \Gamma_F}{\delta\Psi_a}, \hspace{0.5cm}
\bar{K}_{ab} = \frac{\delta\Gamma_F}{\delta G_{ab}}\, . 
\ee
We observe that $\Gamma_F$ generates loop diagrams with the full
propagator $G_{ab}$. The factor $\det(G)^{-1/2}$ removes the one-loop
diagram already included in equ.~(\ref{Gam_2PI_BF}). The action $S_3$
is defined in analogy with equ.~(\ref{S_BF})
\be 
\label{S_3}
 S_3[\Psi_a,\delta\psi_a] = S[\Psi_a+\delta\psi_a]
   - \frac{\delta S}{\delta\Psi_A}\, \delta\psi_A
   - \frac{1}{2}\, 
    \frac{\delta^2 S}{\delta\Psi_A\delta\Psi_B}\, 
           \delta\psi_A\delta\psi_B\, . 
\ee
Note that for a cubic interaction $S_3$ is not a function of the 
background field $\Psi_a$, and the only vertex in $S_3$ is that 
of the original action, given by equ.~(\ref{S_I_K}) with $\psi_a
\to\delta\psi_a$. If we include a quartic interaction term, then 
$S_3$ contains a new vertex of the 
form 
\be 
 S_3[\Psi_a,\delta\psi_a] 
    = \frac{\kappa\lambda_4}{3!}\int d^3x\, dt\, d_{abcd}
  \left[ (\nabla^2\Psi_a)\delta\psi_b\delta\psi_c\delta\psi_d
+ 3 (\nabla^2\delta\psi_a) \Psi_b\delta\psi_c\delta\psi_d \right]\, , 
\ee
where $d_{1222}=1$ and all others $d_{abcd}=0$. Finally, we note that 
the source terms in equ.~(\ref{Gam_2PI_sources}) remove tadpoles 
order by order in the loop expansion of equ.~(\ref{Gam_2PI_int}).

\begin{figure}[t]
\begin{center}
\includegraphics[width=0.85\textwidth]{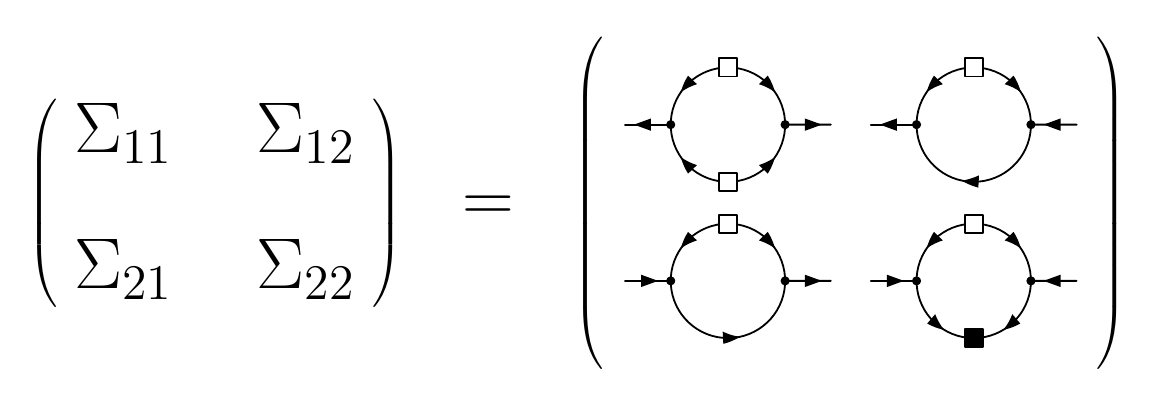}
\end{center}
\caption{\label{fig:2PI-DS}
Schwinger-Dyson equation for the self energies in the 2PI formalism. }
\end{figure}

The Legendre transform in equ.~(\ref{Gam_2PI}) is known as the 2PI
effective action, because the loop expansion of equ.~(\ref{Gam_2PI_int})
corresponds to the sum of two-particle irreducible diagrams. The leading 
two-loop diagrams are shown in Fig.~\ref{fig:2PI-loop}. These diagrams
give
\bea
 \Gamma_F[G_{ab}] &=& \frac{1}{2}\left( \frac{\kappa\lambda_3}{2}\right)^2
   \int d^3x\,dt\, d^3x'\,dt'\, \bigg\{ 2
     \Big( \nabla_x^2\nabla_{x'}^2 G_{11}(x,t;x',t')\Big)
         \Big( G_{22}(x,t;x',t')\Big)^2 
          \nonumber \\
  && \label{Gam2_2loop} \hspace{0.4cm}\mbox{}
    + 4 \Big( \nabla_x^2 G_{12}(x,t;x',t') \Big)
        \Big( \nabla_{x'}^2 G_{21}(x,t;x',t') \Big)
        G_{22}(x,t;x',t')
    \bigg\}\, . 
\eea
Note that, as explained above, this result only depends on $G_{ab}$, 
and not on $\Psi_a$. We can now study the equation of motion that 
follows from equ.~(\ref{Gam_2PI_BF}). We find
\be
\label{Sigma_2PI}
\Sigma_{ab} \equiv [G^{-1}]_{ab} - [G_0^{-1}]_{ab}
 = 2\, \frac{\delta\Gamma_F[G]}{\delta G_{ab}} \, . 
\ee
This is a self-consistent equation for the matrix self-energy
$\Sigma_{ab}$. Note that in equ.~(\ref{Sigma_1PI},\ref{D_1PI}) we 
denoted $\Sigma=\Sigma_{12}$ and $\delta D = \Sigma_{11}$. The 
equation of motion for $\Psi_a$ is the same as in equ.~(\ref{Psi_EoM}),
where $\Sigma$ is given by the solution of equ.~(\ref{Sigma_2PI}).
For translationally invariant systems the consistency equation
(\ref{Sigma_2PI}) is most easily stated in $(\omega,k)$ space. 
Using equ.~(\ref{Gam2_2loop}) we get
\bea
\label{gap_1}
\Sigma(\omega,k) & \equiv & \Sigma_{12}(\omega,k) =  
 (\kappa\lambda_3)^2 \int d^3k'd\omega'
   (k+k')^2 k^2 G_{22}(\omega',k') 
     G_{21}(\omega+\omega',k+k') \, ,  \\
\label{gap_2}     
\delta D(\omega,k) & \equiv & \Sigma_{11}(\omega,k) =  
 \frac{(\kappa\lambda_3)^2}{2} \int d^3k'd\omega'
   k^4 G_{22}(\omega',k') 
     G_{22}(\omega+\omega',k+k') \, , 
\eea
where $G_{ab}$ is self-consistently determined by 
equ.~({\ref{Sigma_2PI}). In components this relation is given 
in equ.~(3.4,3.6) in reference \cite{Chao:2020kcf}. For $G_{ab}=
[G_0]_{ab}$ we get the perturbative result in 
equ.~(\ref{Sigma_1PI},\ref{D_1PI}). Note that at the stationary 
point $G_{11}$ and $\Sigma_{22}$ vanish. This also implies that 
at the stationary point the 2-loop approximation to $\Gamma_F$
in the cubic theory vanishes.
 
\section{Gap equation in mixed representation}                    
\label{sec_mixed}

\begin{figure}[t]
\includegraphics[width=0.49\textwidth]{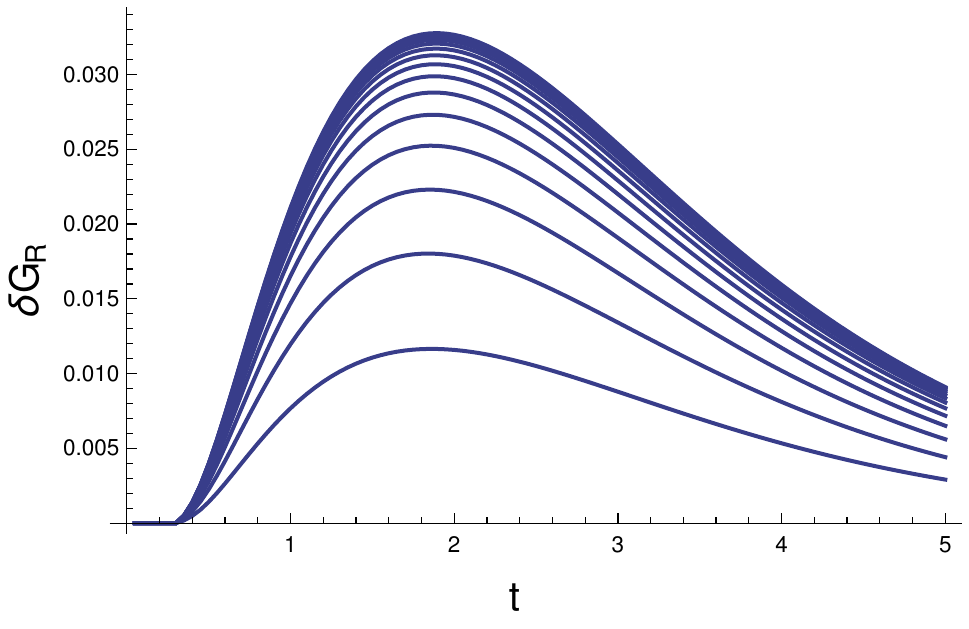}
\includegraphics[width=0.49\textwidth]{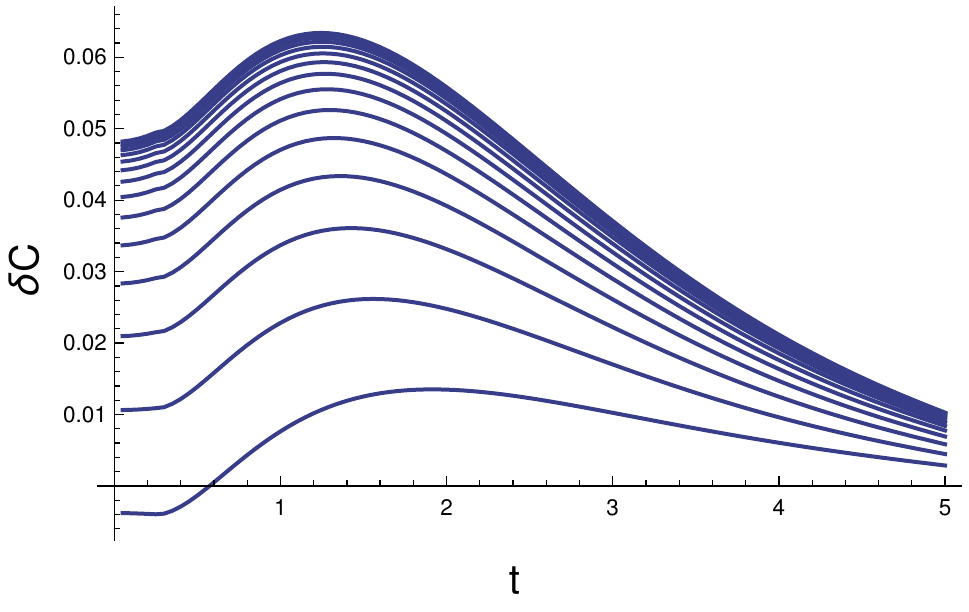}
\caption{Loop corrections to the retarded Green function $\delta
G_R(k,t)=G_R(k,t)-G_R^0(k,T)$ (left panel) and correlation function
$\delta C(k,t)=C(k,t)-C^0(k,t)$ (right panel) as a function of $t$
for fixed $k=1$. We have also chosen $\lambda_3=2.5$ and $D=1$. 
The different curves show the convergence of fixed-point iterations
of the Schwinger-Dyson equation. The lowest curve is the result
of the first iteration, corresponding to the one-loop result.}  
\label{fig:num}
\end{figure} 

 A practical approach to implementing equ.~(\ref{gap_1}-\ref{gap_2}) 
is to solve the integral equations in a mixed representation 
$\Sigma_{ab}(t,k^2)$. In the mixed representation
\bea 
\label{gap_1_m}
\Sigma(t,k^2) &=& (\kappa\lambda_3)^2  \int d^3k'\, k^2(k+k')^2 \,
  C(t,k') \, G_R(t,k+k') \, , \\
\label{gap_2_m}  
\delta D(t,k^2) &=& \frac{(\kappa\lambda_3)^2}{2}  \int d^3k' \,  
  k^4 C(t,k') \, C(t,k+k') \, . 
\eea 
In the mixed representation we need to solve the Dyson equation 
to close this set of equations. We have
\be 
\label{Dyson_equ}
G_{ab}(t,k^2) = G^0_{ab}(t,k^2) - \int dt_1 \int dt_2\, 
  G^0_{ac}(t_1,k^2)\Sigma_{cd}(t_2-t_1,k^2)G_{db}(t-t_2,k^2)\, .  
\ee
The matrix product can be decomposed in terms of retarded, 
advanced, and symmetric functions. We find
\bea 
\label{Dyson_R}
G_R(t,k^2) &=& G_R^0(t,k^2) - \int dt_1\, \int dt_2 \,
  G_R^0(t_1,k^2)\Sigma_R(t_2-t_1,k^2)G_R(t-t_2,k^2), \\
\label{Dyson_S}
C(t,k^2)  &=&  C^0(t,k^2) -   \int dt_1\, \int dt_2 \,
 \Big[   G_R^0(t_1,k^2)\delta D(t_2-t_1)G_A(t-t_2,k^2) \nonumber \\
     && \hspace{4.2cm}\mbox{}
   + G_R^0(t_1,k^2)\Sigma_R(t_2-t_1) C(t-t_2,k^2) \nonumber \\
     && \hspace{4.2cm}\mbox{}
   + C^0(t_1,k^2)\Sigma_A(t_2-t_1)G_A(t-t_2,k^2) 
\Big],
\eea
where $C=G_{22}$, $G_R=G_{21}$, $G_A=G_{12}$ as well as $\Sigma_R
\equiv \Sigma = \Sigma_{12}$, $\Sigma_A=\Sigma_{21}$, $\delta D=
\Sigma_{11}$. The structure of equ.~(\ref{Dyson_R},\ref{Dyson_S})
ensures that the correlation functions have the correct symmetry,
$G_R(t,k^2)=G_A(-t,k^2)$ and $C(t,k^2)=C(-t,k^2)$. The free propagator 
in the mixed representation given by
\be 
\label{G_0_m}
G^0_{ab}(t,k^2) = \left( \begin{array}{cc} 
   0 &   \Theta(-t)\, e^{Dtk^2} \\
 \Theta(t)\, e^{-tDk^2} & \frac{T}{m^2}\, e^{-D|t|k^2}
\end{array} \right) \, . 
\ee
\begin{figure}[t]
\begin{center}
\includegraphics[width=0.66\textwidth]{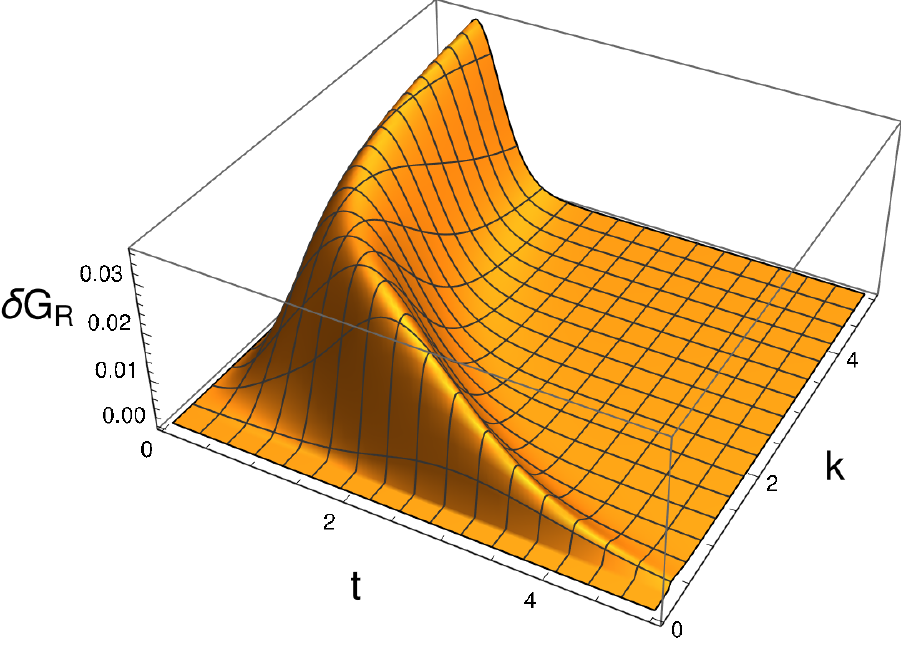}
\end{center}
\caption{Loop corrections to the retarded Green function $\delta
G_R(k,t)$ as a function of $t$ and $k$. Parameters as in 
Fig.~\ref{fig:num}.}  
\label{fig:num3d}
\end{figure} 
In the mixed representation, the gap equation 
(\ref{gap_1_m},\ref{gap_2_m}) is UV finite, but there are short-time
singularities in the Dyson equ.~(\ref{Dyson_equ}). This is clear 
from equ.~(\ref{sigR_t_k}), which shows that the one-loop self-energy
contains terms of order $t^{-5/2}$ and $t^{-3/2}$. To 
regularize these singularities we have employed the Pauli-Villars
method, see App.~\ref{sec_PV}. We add a number of hydrodynamic
fields $\chi_i$ with diffusion constants $D_i\gg D$. These fields
do not modify the Green functions for a large time, but the 
coupling constants can be adjusted to remove short-time 
singularities. Changing the $D_i$ while adjusting the couplings
to remove singularities in $\Sigma_{ab}(t,k^2)$ corresponds
to adjusting polynomial terms in $\Sigma_{ab}(\omega,k^2)$.

 A numerical solution of the Schwinger-Dyson 
equations~(\ref{gap_1_m},\ref{gap_2_m}) and (\ref{Dyson_R},\ref{Dyson_S})
is shown in Figs.~\ref{fig:num} and \ref{fig:num3d}. We plot the loop
corrections $\delta G_R=G_R-G_R^0$ and $\delta C=C-C^0$ to the retarded
Green function and the correlation function. Note that the unit of 
length is given by the bare correlation length $\xi=m^{-1}$, and the 
unit of time is given by the relaxation time $\tau=\kappa/\xi^4$. The 
solutions are shown in dimensionless time units $t$ and wave 
number $k$. Numerical solutions are obtained by iterating the gap
equations (\ref{gap_1_m},\ref{gap_2_m}) and the Dyson series
(\ref{Dyson_R},\ref{Dyson_S}) starting with the initial condition
$G_{R,A}=G_{R,A}^0$ and $C=C^0$. This means that after one iteration 
we obtain the one-loop self-energy given in equ.~(\ref{sigR_t_k}). 

 The short time behavior is regularized with the help of two 
Pauli-Villars fields, see Appendix \ref{sec_PV}. For the results 
shown in Figs.~\ref{fig:num} and \ref{fig:num3d} we have used
$(\alpha_{1,2})=(4,5)$, which means that the relaxation time of 
the Pauli-Villars diffusion is four and five times shorter,
respectively, than that of the physical diffusion. There is a 
remaining $t^{-1/2}$ singularity, see App.~\ref{sec_PV}. This
singularity is integrable, but in order to avoid numerical 
difficulties we also impose an explicit short-time cutoff 
$t_c=0.2$. The results in Figs.~\ref{fig:num} and \ref{fig:num3d} 
are obtained for $D=1$ and $\lambda_3=2.5$.

\begin{figure}[t]
\includegraphics[width=0.49\textwidth]{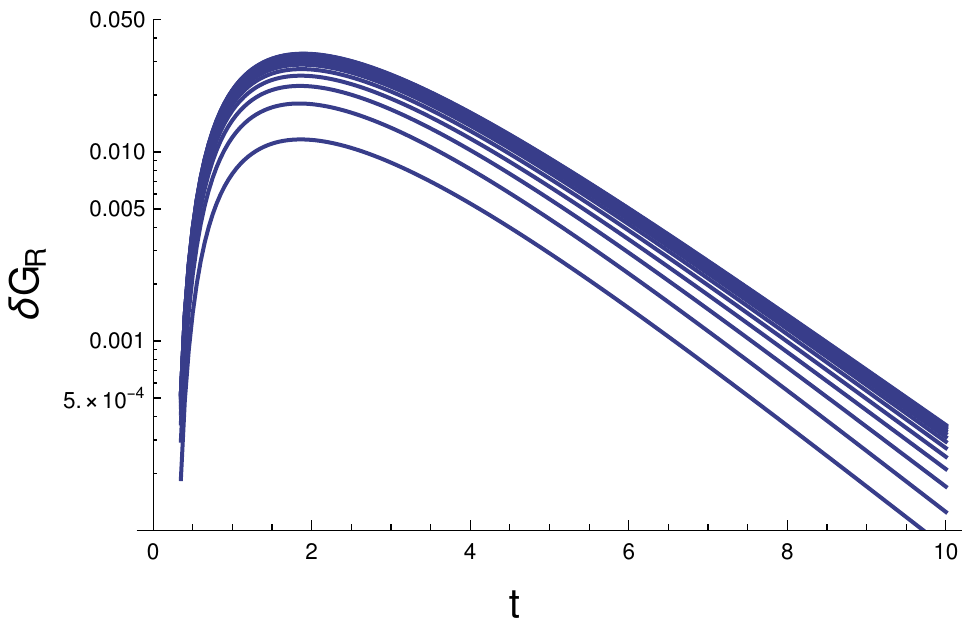}
\includegraphics[width=0.49\textwidth]{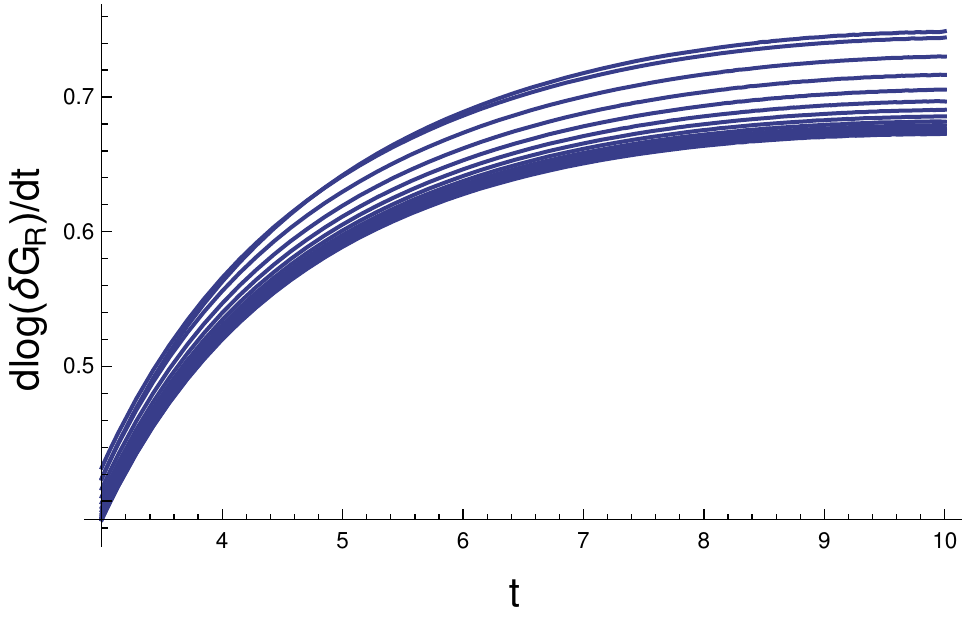}
\caption{Long-time behavior of the loop corrections to the retarded 
Green function $\delta G_R(k,t)$. Parameters are chosen as in 
Fig.~\ref{fig:num}. The left panel shows a logarithmic plot 
of $\delta G_R(t,k^2)$ for $k=1$ up to $t=10$. The bottom curve
is the one-loop result, and the top curve shows the converged 
solution of the Schwinger-Dyson equation. The right panel shows 
the logarithmic derivative of $\delta G_R(t,k^2)$ with respect 
to $t$. The top curve is the one-loop result, and the bottom
one is the converged solution. }  
\label{fig:num-log}
\end{figure} 

 We observe that the iterative solution of the Schwinger-Dyson
equation is indeed convergent. For small $\lambda_3\sim 1$ the 
final solution is very close to the one-loop result. For larger 
values of $\lambda_3$ we observe significant corrections. In 
the regime $t\lsim 1$ these deviations are very sensitive to 
the form of the regulator, but for $t\gsim 1$ the corrections
are universal in the sense that changes in the regulator can 
be compensated by changes in the coupling. Loop corrections 
renormalize the strength of the coupling $\lambda_3$. There is 
a critical value of the bare coupling beyond which no solutions 
of the Schwinger-Dyson can be found. The value of the critical 
coupling depends on the choice of the regulator. For the parameters used
in Figures, $\alpha_{1,2}=(4.0,5.0)$, the value of the critical 
coupling is $\lambda_{3,c}\simeq 2.58$. In Fig.~\ref{fig:num-log}
we analyze the long-time behavior of loop corrections to the 
retarded Green function in more detail. The left panel shows
a logarithmic plot of $\delta G_R(t,k)$ up to larger times
$t\leq 10$, and the right panel shows the logarithmic derivative
of $\delta G_R(t,k^2)$ with respect to $t$. The one-loop correction
decays as $\exp(-Dk^2t/2)$. Delacr{\'e}taz noticed that $n$-loop terms
scale as $n!\exp(-Dk^2t/n)$, and conjectured that the long-time
behavior of the diffusion cascade is $\exp(-\alpha\sqrt{DK^2t})$
with $\alpha\sim 1$ \cite{Delacretaz:2020nit}. Our results are 
consistent with the emergence of a cascade -- we observe that the 
logarithmic derivative of $\delta G_R$ decreases as higher and
higher loops are summed -- but it is difficult to establish 
the behavior at asymptotically long times. 

\section{Conclusions and Outlook}

 In this work we have studied the 1PI and 2PI effective actions
for the stochastic diffusion equation. We have numerically 
investigated solutions of the Schwinger-Dyson equation derived
from the 2-loop 2PI action in a model with a cubic coupling. 
We find that higher loop corrections summed by the Schwinger-Dyson
equation renormalize the coupling constant, and we observe 
indications of a diffusion cascade at long times. 

 This existence of the diffusion cascade implies that the long-time
behavior of diffusion is non-perturbative, even in a non-critical
fluid. We can estimate the relevant time scale based on 
equ.~(\ref{sigR_t_k}). For this purpose we use $m=\xi^{-1}$
and $\kappa=\xi^4/\tau$, where $\xi$ is the correlation length
and $\tau$ is the relaxation time. We also write the non-linear
coupling as $\lambda_3=g_3/(\xi^{3/2}T^{1/2})$, where $g_3$ is 
dimensionless. The one-loop correction $\delta G_R\sim
\exp(-Dk^2t/2)$ decays more slowly than the tree level term
$G_R^0=\exp(-Dk^2t)$. The two terms are comparable if 
\be
 t \gsim \frac{2\tau}{(k\xi)^2} \log\left(
    \frac{2}{\alpha g_3^2 (k\xi)^2} \right)\, ,
\ee
where $\alpha=1/(64\sqrt{2}\pi^{3/2})$ is the numerical constant
in equ.~(\ref{sigR_t_k}). For $g_3=O(1)$ fluctuations with wave
number $k=\xi^{-1}$ are non-perturbative for $t\gsim 2\tau\log(2
/\alpha)\sim 12 \tau$. In the case of a relativistic heavy 
ion collision we have previously used the estimate $\xi\simeq 1.2$ 
fm and $\tau=1.8$ fm \cite{Akamatsu:2018vjr,Martinez:2019bsn}.
This means that non-perturbative effects set in for $t\gsim 20$
fm, too large to be relevant in a heavy ion collission in which
the life time of the fireball is on the order of 10 fm. Modes
with $k>\xi^{-1}$ become non-perturbative earlier, but in that
case higher order hydrodynamic effects are also important. 

 The methods described in this work can be extended in a variety
of ways. One direction is to compute the 3PI effective action, 
and determine the self-consistent equation for the three-particle
vertex. Note that the third Legendre transform, even though it is
usually referred to as the 3PI effective action, is not the sum of 
all 3PI diagrams. Indeed, in a theory with only a cubic interaction 
there are no 3PI diagrams, but one can nevertheless construct an
effective action that generates a self-consistent vertex function
\cite{deDominicis:1964zz,deDominicis:1964zzb}.

 Another interesting direction is to consider the critical
regime, both for a purely diffusive theory and for a theory
of a conserved density coupled to the momentum density of the 
fluid. These theories are known as model B and model H in the 
classification of Hohenberg and Halperin \cite{Hohenberg:1977ym}.
Model H describes the critical endpoint in a single component 
fluid, and is also believed to describe a possible endpoint 
of the quark-gluon plasma transition in QCD \cite{Son:2004iv}.
There are several approaches for extracting the critical
correlation functions. The first is the $\epsilon$-expansion, 
in which physical quantities are computed as an expansion around 
the critical dimension. A second approach, known as mode coupling
theory \cite{Kawasaki:1970}, is based on approximate solutions 
of self-consistent equations. The nPI method provides a way to
systematically check these approximations and extend the 
results to higher $n$-point functions. A related approach
is the functional renormalization group (FRG), which has also been
applied to critical dynamics \cite{Canet:2007}. An advantage
of functional methods such as the FRG or the nPI method is 
that they can serve as a starting point for deriving
approximate kinetic equations. These kinetic equations may 
provide a practical approach to critical dynamics in systems, 
such as heavy ion collisions, in which there is a non-trivial
background flow. 

\acknowledgments
This work is supported by the U.S. Department of Energy Office 
of Nuclear Physics through Contract DE-FG02-03ER41260 (T.S.). 
The work of J.C. is supported by start-up funding from 
Jiangxi Normal University under grant No. 12021211. 
T.S. would like to thank Alexander Kemper for useful discussions, 
and Luca Delacr{\'e}taz for pointing us to \cite{Delacretaz:2020nit}.

\appendix
\section{Pauli-Villars Regulator}
\label{sec_PV}

 We consider the effective lagrangian for a diffusive field
$\psi$ in the presence of Pauli-Villars regulator fields $\chi_i$
($i=1,\ldots,N_{\it PV}$),
\be
\label{L_PV}
{\cal L} = \tilde\psi\left( \partial_t\psi - \kappa \nabla^2
  \frac{\delta {\cal F}}{\delta\psi} \right) 
   + \tilde\psi \kappa T \nabla^2 \tilde\psi + 
    \tilde\chi_i \left( Z_{ij}^{-1}  \partial_t \chi_i
      -\bar{\kappa}_{ij}\nabla^2 \frac{\delta{\cal F}}{\delta\chi_j}\right)
   + \tilde\chi_i\bar\kappa_{ij}T\nabla^2 \tilde\chi_j \, , 
\ee
with 
\be
{\cal F} = \int d^3x\, \Big\{ \frac{1}{2}(\nabla\psi)^2
  + \frac{1}{2} m^2\psi^2 + \frac{\lambda_3}{3!}\psi^3
  + \frac{1}{2}(\nabla\chi_i)^2
  + \frac{1}{2} \bar{m}_{ij}^2\psi_i\psi_j 
    + \frac{\bar{\lambda}_{ij}}{2}\psi\chi_i\chi_j
 \Big\}\, . 
\ee
In the following we will assume that $Z_{ij}=\delta_{ij}$. The 
lagrangian in equ.~(\ref{L_PV}) is of the form
\be 
 {\cal L} = \tilde{s}_a\left( \partial_t s_a - V(s) \right) 
   +  \tilde{s}_a \kappa_{ab}  \tilde s_b\, , 
   \hspace{1cm}
   V_a(s) = -\kappa_{ab}\frac{\delta {\cal F}}{\delta s_b}\, ,
\ee
where $\kappa_{ab}$ is a symmetric matrix and $s_a=(\psi,\chi_i)$
as well as $\tilde{s}_a=(\tilde\psi,\tilde\chi_i)$. This lagrangian 
has a time reversal invariance 
\bea 
{\cal T} s_a(t) & \to & \;\;\;\; s_a(-t) \, ,  \\
{\cal T}\tilde{s}_a (t) & \to &  \left[ 
 - \tilde{s}_a(-t) + \frac{\delta {\cal F}}{\delta s_a}\right]\, . 
\eea
under which ${\cal L}\to {\cal L} + (d{\cal F})/(dt)$. The time
reversal symmetry ensures that detailed balance and 
fluctuation-dissipation relations are satisfied in the presence 
of the regulator fields. 

 We will also take the matrices $\bar{m}_{ij},\bar{\kappa}_{ij}$
and $\bar{\lambda}_{ij}$ to be diagonal, and denote $\bar{m}_{ij}
=\delta_{ij}\bar{m}_i$ etc. The quadratic part of the lagrangian is
\be
{\cal L} = 
 \tilde\psi\left( \partial_t -\kappa m^2\nabla^2\right)\psi
 +\tilde\psi \kappa T\nabla^2 \tilde\psi
 + \tilde\chi_i\left( \partial_t 
          - \bar{\kappa}_i^2 \bar{m}_i^2\nabla^2\right)\chi_i
 +\tilde\chi_i \bar{\kappa}_i T \nabla^2 \tilde\chi_i\, . 
\ee
We allow the $(\tilde\chi_i,\chi_i)$ to be ghost fields, so that 
loops acquire an extra minus sign. We also define $\bar{\kappa}_i=
\alpha_i\bar\kappa$ and set $\bar{m}^2_i=m^2$. This implies 
that the diffusion constant of the Pauli-Villars fields is 
\be
 \bar{D}_i = \alpha_i D \, . 
\ee 
The non-linear interaction terms are
\be 
{\cal L} = -\frac{\kappa\lambda_3}{2} 
  \left(\nabla^2\tilde\psi\right) \psi^2
 - \frac{\kappa\bar{\lambda}_i}{2}
  \left(\nabla^2\tilde\psi \right) \chi_i\chi_i
 - \bar\kappa\bar{\lambda}_i\, 
  \left(\nabla^2\tilde\chi_i \right) \psi\chi_i\, , 
\ee
and we define
\be
\bar\lambda_i^2 = c_i \lambda_3^2\, . 
\ee
The Pauli-Villars fields are then characterized by the 
parameters $(\alpha_i,c_i)$. We will adjust these parameters
to remove the UV divergences in the self energy.

 Consider the one-loop contribution to the retarded 
self energy $\Sigma(t,k^2)$, see equ.~(\ref{sigR_t_k}). 
In the presence of Pauli-Villars fields the leading short 
time behavior is
\be
\Sigma(t,k^2) =  \frac{3}{64\sqrt{2}\pi^{3/2}}
\frac{\lambda_3^2T}{m^6 \sqrt{D}}
  \left\{ \frac{1}{t^{5/2}}
      \left[ 1 +\sum \frac{c_i}{\alpha_i^{1/2}} \right]
- \frac{Dk^2}{6t^{3/2}} 
     \left[ 1 + \sum_i c_i\alpha_i^{1/2} \right]
    + O\left(\frac{1}{t^{1/2}}\right)
 \right\}. 
\ee
The terms of order $t^{-5/2}$ and $t^{-3/2}$ lead to divergences 
in the Fourier transform $\Sigma(\omega,k^2)$ and the convolution
integral in the Dyson equation, see equ.~(\ref{Dyson_equ}). If 
we define a short-time cutoff $t_c\sim (D\Lambda^2)^{-1}$ then 
$\int dt/t^{5/2}\sim \Lambda^3$ and $\int dt/t^{3/2}\sim \Lambda$. 
We can eliminate these divergences by choosing suitable 
Pauli-Villars fields. These have to satisfy
\be
1 + \sum \frac{c_i}{\alpha_i^{1/2}} = 0\, , \hspace{1cm}
1 + \sum_i c_i\alpha_i^{1/2}  = 0 \, . 
\ee
The minimal number of fields is two. In this case we have
\be 
\label{PV_paras}
c_1 = \alpha_1^{1/2}  \frac{\alpha_2-1}{\alpha_1-\alpha_2}\, ,
\hspace{1cm}
c_2 = - \alpha_2^{1/2} \frac{\alpha_1-1}{\alpha_1-\alpha_2}\, .
\ee
We can choose any set of $\alpha_{1,2} > 1$ as long as $\alpha_1
\neq \alpha_2$. Different choices of these parameters correspond
to different values of higher order transport coefficients.
We note that one of the $c_i$ is negative, corresponding to 
a ghost field. A consistency check is provided by computing
$\delta D(t,k^2)$. Once the $(\alpha_i,c_i)$ are fixed by 
the requirement that short-time singularities in $\Sigma(t,k^2)$
are removed, then $\delta D(t,k^2)$ should be non-singular
as well. This is indeed the case.

\bibliography{bib}

\end{document}